# Plasma Experiments with Relevance for Complexity Science

**Erzilia Lozneanu, Sebastian Popescu** and **Mircea Sanduloviciu**
Plasma Physics Department, 'Al.I.Cuza' University
6600 Iasi, Romania
msandu@uaic.ro

The goal of this paper is the identification of the physical processes at the origin of the nonlinear behavior of a plasma conductor when an external constraint gradually departs the system from thermal equilibrium. This reveals the presence of a self-organization scenario whose final product depends on the magnitude of the applied constraint. At first it appears a complexity whose stability is ensured by the presence of an electrical double layer. By increasing the external constraint the complexity transits into an autonomous state whose existence is related to a rhythmic exchange of matter and energy with the surrounding environment, sustained and controlled by a proper dynamics of the double layer. The results are potentially important for developing a general strategy of nonequilibrium physics, suggesting answers to challenging problems concerning the mechanism that could explain the appearance of self-organized complexities in laboratory and nature.

## 1. Introduction

Plasma Physics is potentially a Physics of Complexity in which self-organization phenomena can be frequently observed [1,2]. In collisional plasma such phenomena usually appear as spatial and spatiotemporal patterns observable as beautiful colored, stationary or moving, space charge configurations. Their appearance can be initiated in two different ways [2]. First, by gradually increasing of a local gradient of the kinetic energy of electrons under controllable laboratory conditions (i.e. intermittent self-organization). Second, by creating a well localized hot plasma in nonequilibrium, by sudden injection of matter and energy, and its natural relaxation towards a self-organized complex structure (i.e. cascading self-organization scenario). The final product of both self-organization scenarios is a complexity able to ensure its own existence by a rhythmic exchange of matter and energy with the surrounding environment.

## 2. Intermittent self-organization scenario

By investigating the causes that are at the origin of the nonlinear behavior of a gaseous conductor we revealed the genuine physical basis of pattern formation. It is charge accumulation in two adjacent opposite net electrical space charges, determined by the spatial separation of the excitation and ionization cross section functions [3-5]. Such phenomena appear in plasma when a sufficiently strong gradient of electrons' kinetic energy is locally created.

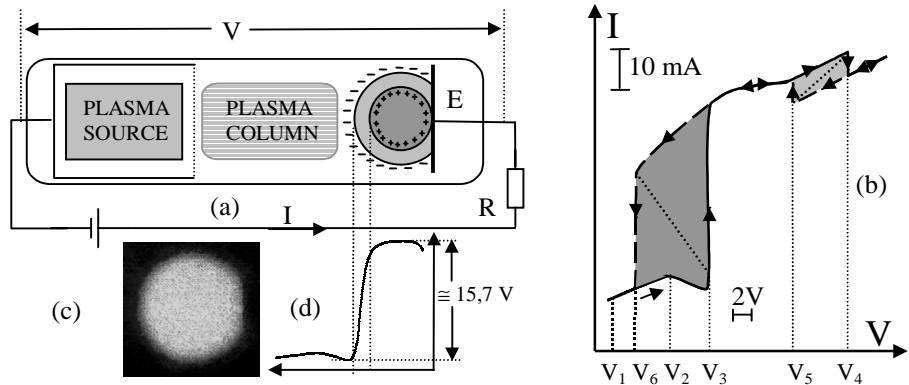

**Figure 1.** Schematic representation of the Ar plasma diode and the internal space charge configuration of the fireball (a), photograph of the fireball (b), potential drop on the double layer bordering the fireball (c), and static I(V) characteristic of the plasma diode (d).

In our experimental device [Fig.1(a)] we have created a gradient of kinetic energy of electrons by their local acceleration towards the positively biased electrode E. The nonlinear variation of the plasma conductivity, related to the magnitude of V is emphasized in the static I(V)-characteristic shown in Fig. 1(b). By identifying the causes that produce the abruptly changes of I for critical values of V (marked by subscripts), it becomes possible to reveal the succession of the sequences of a physical scenario that explains the genesis of a complexity formed by an intermittent self-organization scenario when V is increased. As we will show, two important sequences of this scenario are revealed in Fig 1(b). They appear for the critical values $V_3$ and $V_4$. Thus, the first one indicates the amount of matter and energy that must be injected into the system for the spontaneously self-assemblage of a stable self-organized complexity, known as fireball [4], bordered by an electrical double layer (DL) (spatial pattern). The second one indicates the amount of matter and energy additionally required for "animating" the fireball. This means the spontaneous transition of the fireball into an autonomous state in which it is able to ensure its own existence by a rhythmic exchange of matter and energy with the environment, emphasized by periodical limitations of I [Fig. 2(a)]. The exchange process is driven and controlled by a self-sustained dynamics of the DL (spatiotemporal pattern). Decreasing V we remark the presence of hysteresis phenomena that explains the ability of the plasma conductor to work as a generator of oscillations relates to its S-shaped, respectively Z-shaped bistability behavior.

By applying a gradual increasing V on the Ar plasma column, this is driven away from equilibrium. The deviation from equilibrium is felt in the region from the front of E, where the electric field penetrates the plasma, so that the low energetic (thermalized) plasma electrons are accelerated towards E. When V reaches the value of the neutrals' excitation potential ($V_2$) a part of the accelerated electrons lose their momentum and, consequently, are accumulated. The maximum concentration of the net negative space charge, formed in this way, is located at a certain distance from E where the neutrals' excitation cross section function suddenly increases [3-5]. Its position depends on V. The condition for the maintenance of the net negative space charge is ensured by the presence of a dissipative process emphasized by light emission by those atoms that return in their ground state after excitation. The concentration of neutrals is maintained constant by a local statistical equilibrium between excitation and de-excitation processes. The dissipated energy has its origin in ordered kinetic energy of the accelerated electrons that have produced neutral excitations. A barrier effect for I appears simultaneously with the net negative space charge accumulation and its development when V increases explains the decreasing of I in the voltage range $V_2 - V_3$ [Fig 1 (b)].

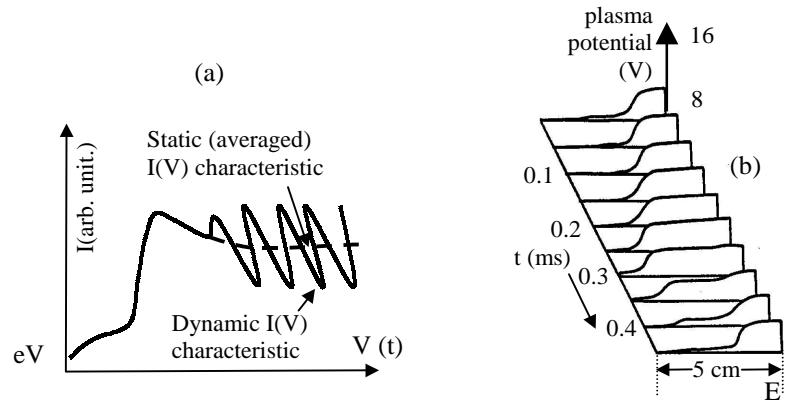

**Figure 2.** (a) Dynamic I(V) characteristic, and (b) space-time picture proving the detachment, motion and disruption of the double layer from the fireball's border.

When V is further increased reaching/exceeding the ionization potential of the gas, the electrons that have not excited neutrals produce ionizations. Since those electrons, as well as those resulting in such processes are rapidly collected by E, a plasma enriched in positive ions appears in front of E. Its presence is the premise for the development of electrostatic forces whose magnitude, related to the concentration of the net opposite space charges, increases simultaneously with V. Acting as long-range correlations, these forces govern, in an increasing manner, the spatial arrangement of electrons and positive ions in front of E, determining the self-assemblage of a planar DL. A critical situation (instability point) appears when $V=V_3$. Here, any small increase of the net positive space charge starts a positive feedback (self-enhancement) mechanism for the positive ion production. Such a mechanism becomes possible because every increase of the net positive space charge implies an increase of the electric field, that, at its turn, determines an increase of electrons' kinetic energy and, considering the energy dependence of the ionization cross section function, also an increase of the net positive space charge. This self-enhancement of the production of positive ions creates a gaseous anode (the high potential side of the double layer) with a rapidly increasing potential. Consequently, the region where electrons are accumulated, after neutral excitations, is shifted away from E. Therefore the initial planar DL transits spontaneously into a state characterized by a local minimum of the free energy. So the fireball appears in front of E when $V = V_3$ [Fig 1 (c)]. Investigating the internal structure of the fireball, we discovered the presence of an ordered arrangement of the electrical space charges, including the DL that protects, like a cell membrane, the complexity from the surrounding plasma [Fig 1(a) and (d)]. One of the most important results of our experiments is the fact that, after the fireball's appearance, the role, initially played by E, is "assumed" by the nearly spherical DL. The spontaneous self-assemblage process of the fireball evidences, without any exaggeration, the "birth" of a complexity through self-organization. Its "vitality" is proved by the hysteresis branch $V_3 – V_6$ that emphasizes a special kind of "memory " typical to all self-organized complexities that are able to preserve their identity also for conditions less favorable than those required for their birth. Evidently the spontaneous appearance of the fireball is accompanied by sudden entropy "expulsion". Therefore, during the very short time sequence, corresponding to the fireball birth, the system defies the second principle of the thermodynamics. After birth, the fireball acts as an "active" complexity ensuring by itself the local acceleration of electrons required for its existence. In this state the DL from the fireball's border works as an additional source of charged particles because the electrons accelerated in its potential drop produce ionizations. We note that the sudden increase of I when $V = V_3$ is not the cause of the fireball genesis but, contrary, the genesis of the fireball is at the origin of I increasing. The stable phase of the fireball in the voltage range $V_3 – V_4$ requires a continuous feeding process (transport of electrons), ensured by the external dc power supply.

   The stability of the fireball is possible as long as the DL ensures its consistence. In this confined state the DL acts as a "membrane" that could be characterized by a certain "surface tension" that must compensate the kinetic pressure of the neutrals, positive ions and electrons within the volume bordered by it. Increasing V, the kinetic pressure inside the fireball reaches a critical value for which the fireball "explodes" [4]. In this moment the DL detaches from the fireball's border, transiting into a moving phase. This transition, taking place when $V = V_4$, crucially changes the self-assemblage mechanism of the DL. Thus, running through the plasma column, the DL becomes able, by self-adjusting its velocity, to ensure a part of the "food" (electron flux) required for its self-maintenance. The other part of the electron flux needed for the DL maintenance is ensured by I. Evidently, the self-maintenance of the DL in a moving phase becomes possible only when the "food" present in the plasma (neutrals and electrons) ensures the required rates of neutrals' excitation and ionization. As is known, the excitation and ionization rates depend (besides on the intensity of the electron flux) on the potential drop over the DL where the electrons are accelerated.

This potential drop is maintained by the self-enhancement mechanism above described by which the production of positive ions is self-adjusted so that all conditions required for the maintenance of the DL are ensured. Because of the lateral limitation of the plasma column the DL appears in our experimental device [Fig 1(a)] as a moving striation that propagates away from E [Fig. 2 (b)]. The departure of the DL from E creates the conditions for the self-assemblage of a new DL in the region where the first one was created. Therefore I depends on the barrier present at the negative side of the expanding DL but also on the barrier formed in front of the new DL. This makes possible an internal triggering mechanism related to the fact that the development of the barrier in front of the new DL diminishes I at a value for which the electron flux traversing the moving DL is not sufficient to maintain the moving DL existence. In this moment it disrupts, releasing the electrons and positive ions from its structure. Moving as a bunch towards E, the electrons reach the region where the new DL is in a developing phase. Traversing it, the rates of excitation and ionization increase to values for which the new DL again detaches from the fireball's border. In this way the DLs generated periodically in front of E run successively through a certain part of the plasma conductor. This DL dynamics explains the periodical limitation of I shown in Fig. 2(a). The sudden decrease of I [Fig. 1(b)], that accompanies the appearance of the moving DL, emphasizes that the energy dissipated becomes smaller, proving the transition into a state for which the production of entropy becomes minimal. The hysteresis loop $V_4$- $V_5$ reveals a new quality of the fireball, namely the ability to ensure its existence by a rhythmic exchange of matter and energy with the surrounding plasma also for matter and energy injection less than those that have determined its transition into the dynamical regime.

We notice that the self-maintenance of the spatiotemporal pattern involves the generation of moving space charge domains in front of the plasma source (cathode) that move towards E [6]. These appear because every disruption of a DL generated in front of E is accompanied by the quickly collection of the electrons present in the plasma column by E. As a consequence, in front of the plasma source the plasma potential suddenly increases at a value close to the voltage of E. In this way the region where electrons are accumulated after neutrals' excitation is shifted in front of the plasma source. So, in this region a negative potential barrier develops. In its further evolution the barrier "moves" as a "space-charge domain" towards E. Arriving at the DL, generated after self-organization in front of E, it determines an increasing of the electron flux at a value sufficient to start the DL moving phase. After running a certain part of the plasma column it disrupts determining the sudden increase of the plasma column potential in front of the plasma source. In this way an internal positive feedback mechanism ensures the periodicity of the phenomenon.

The described DL dynamics reveals a new insight into the mechanism by which the plasma column works as **n**egative **d**ifferential **r**esistance (NDR). The N-shaped NDR is associated with the Z-shaped bistability shown in Fig. 1(b). Its physical basis is the dynamics of the complexity formed in front of the anode that also triggers the formation of moving domains at the cathode and their disruption at the anode. Such phenomena were observed in a Gunn diode connected to a resonant cavity, but they appear also in a plasma diode when a part of the plasma conductor becomes able to support ion-acoustic oscillations [6]. These similarities suggest a common physical basis at the origin of the N-shaped NDR: the dynamics of the self-organized complexity formed in front of the anode determines the periodical limitation of I. That means that this self-organized complexity really acts as the "vital" part of both oscillators. A similar conclusion can be drawn considering the S-shaped NDR shown in Fig 1(b), whose appearance in semiconductors is usually related to current filaments. In gaseous conductors these filaments are related to the propagation of the DLs from the border of the complexity, formed in front of the anode, towards the cathode, usually observed in plasmas with negative ions.

We note that because of experimental difficulties, the phenomena present in front of the anode of a semiconductor working as NDR are not known at the present time. Therefore the described plasma experiment proving that the self-organized complexities in front of the anode are the genuine cause of NDR could be elucidative also for problems not yet conclusively solved in solid-state physics [8]. These results justify, in our opinion, the attempt to prove that self-organization is the phenomenon that may provide a paradigmatic shift in science, offering solutions for challenging problems in chemistry [9] and biology [10].

## 3. Cascading self-organization scenario

A self-organization scenario initiated by sudden injection of matter and energy takes place when a relatively hot plasma is locally produced. This can occur when an electrical spark strikes the surface of a positively charged conductor. Under such conditions an amount of the conductor's material is vaporized creating a plasma that contains thermalized electrons and positive ions. Owing to the difference in mobility between the electrons and positive ions, the first ones are rapidly collected by the conductor, so that a nucleus with an excess of positive ions appears. The net positive charge of the nucleus is additionally enhanced by thermal diffusion of the remaining electrons, so that its potential becomes greater than that of the conductor. When the nucleus is surrounded by an environment that contains thermalized electrons these are accelerated towards it, so that the subsequent evolution of the initially hot plasma depends only on the positive potential of the nucleus. Thus, when the electrons accelerated towards the positive nucleus obtain energies sufficient to produce neutrals' excitation and ionization, the hot plasma evolves (relaxes) into a complexity similar to that created by an intermittent self-organization scenario. Because the transition into a state of local minimum of the free energy involves the formation of an enclosed spherical double layer, the complexity detaches from the conductor, appearing as a free-floating fireball. In laboratory such phenomena, known as balls of fire, appear in low voltage arcs, and in nature as lightning balls [5].

The lifetime of the complexity depends on the number and nature of the constituents of the medium that surrounds it. When this medium is plasma-like containing a sufficient number of neutrals and electrons, the further existence of the free floating fireball becomes possible by a rhythmic exchange of matter and energy with the surroundings. This dynamics, sustained and controlled only by internal processes, ensures all the operations required for the fireball's existence: capture and transduction of energy, preferential exchange of matter across the system's boundary and internal transformation of matter in the ongoing synthesis of all system's components.

For explaining such a complex mechanism we start from the premise that, immediately after its birth, the potential drop over the spherical DL is so high that the described production of positive ions (by a self-enhancement mechanism) determines its expansion. During this expansion, a new DL begins to form at the border of the nucleus. Its development decreases the electron flux towards the nucleus determining the disruption of the previously DL and, consequently, the delivery of electrons bounded at its structure. After acceleration towards the nucleus, the electrons traverse the new DL determining an increase of its potential drop at a value for which its expansion can start. Evidently, the stationary state of the fireball, ensured by successive generation and disruption of the DL at its border, requires the presence of a mechanism by which the nucleus is positively recharged periodically. This becomes possible when the electron carried by the current, accelerated within the expanding DL, reaches the border of the nucleus with energies sufficient to start the shelling off process of the new DL, and also to produce a heating process of the nucleus. The heating process is decisive in ensuring, by thermal diffusion, the required "recharging" mechanism of the nucleus at a voltage for which the described DL dynamics can repeat. Since, during the DL expansion and disruption, positive ions are transported from the fireball to the surroundings, the kinetic pressure inside it decreases so that neutrals from the surrounding medium are periodically drawn into ("inhaled" by) the nucleus. By such a mechanism the fireball constantly produces and replaces its own constituents within a boundary made by itself. In our opinion the fireball represents the simplest-possible self-organized system, created under controllable laboratory conditions, able to perform every operation required for ensuring its own existence in an apparently similar manner to those by which living beings ensure their viability.

## 4. Double layer dynamics at the origin of anomalous transport

Because each DL consists from two opposite space charges, bounded by electrostatic forces, its detachment and propagation away from the region where it was generated really represents an anomalous transport of matter and energy. The DL dynamics involves the stochastic detachment of the DL, controlled by deterministic chaos, from the region where it was generated. That takes place at a time scale in the order of KHz. Over this time scale overlaps another time scale in the order of MHz, related to the self-assemblage/disruption of the "fine" structure of the DL. The last is related to the accumulation of electrons that have lost their energy by excitation at different energy levels of the neutrals [11]. Such a dynamical system can explain one of the classic problems of physics, namely the appearance of the 1/f noise. With respect to the running sandpile considered as a model for the so called self-organized criticality concept [12], the DL dynamics is evidently a more appropriate phenomenology for explaining, for example, the turbulence phenomenon observed in fusion devices [13].

## 5. Conclusions

Investigating the origin of the nonlinear behavior of a plasma conductor we have identified the sequential steps of two physical scenarios whose final product is a self-organized complexity. Comparing these results with the already published experimental results concerning pattern formation in other physical [8-14], but also chemical and biological systems [9,10], the described phenomenological model suggests the possibility to develop a general strategy, based on self-organization, concerning the appearance of complex systems in laboratory and nature.